# Four Generations of Control Theory Development ?

*Tai Cheng Yang*

*Department of Engineering, University of Sussex, Brighton BN1 9QT, UK*

*Abstract:* this short article presents an opinion that control system study up to date can be divided into four generations; namely, 1 transfer function based; 2 state-space based; 3 networked control systems; and 4 control in the new AI era.

In our control community, in particular in our teaching, we often use the terms "classical control theory" and "modern control theory". History moves forward. The word "modern" here is not appropriate. Today's modern is future's classical. Nevertheless, behind the ambiguous words they are meaningful terms: "transfer function based" for classical control, and "state-space based" for modern control. Therefore, we call $1^{st}$ generation of control theory "transfer function based", and the $2^{nd}$ generation "state-space based". Many overviews of control have been published in journals and conferences. Among the all we have seen, we believe that reference [1] – a 41 page *Automatica* paper – gives an excellent review. From real-word challenges to technology advance, to control theory development, to applications, it gives a good comprehensive survey of the $1^{st}$ and $2^{nd}$ generation feedback control development, covering a long period of time and a wide range of disciplines.

Around the beginning of this century, control system research was entering a new era. Traditional framework of a control system structure – plant, sensor, actuator and a controller, see Figure 1 − is no longer applicable to some new research: (a) System to be controlled is a network of subsystems, see Figure 2 where the system of Figure 1 is only a subsystem in a networked control system; (b) Many control tasks are achieved by a network of distributed local controllers (agents); and (c) New research topics on control of networked behaviour, including consensus, formation and synchronisation, are clearly beyond the scope of the traditional $1^{st}$ and $2^{nd}$ generation framework, where the control task is often stated as:"…… to design a controller, …… so that the system is stable/robust/optimal". With these new trends some new concepts are emerging, for example consensusability, formationability, computability, etc. Third generation of control theory is motivated by Networked Control Systems (NCSs), hence the name. More importantly, two fundamental concepts – stability and controllability – in the $3^{rd}$ generation of control theory are much different from the $2^{nd}$ generation.

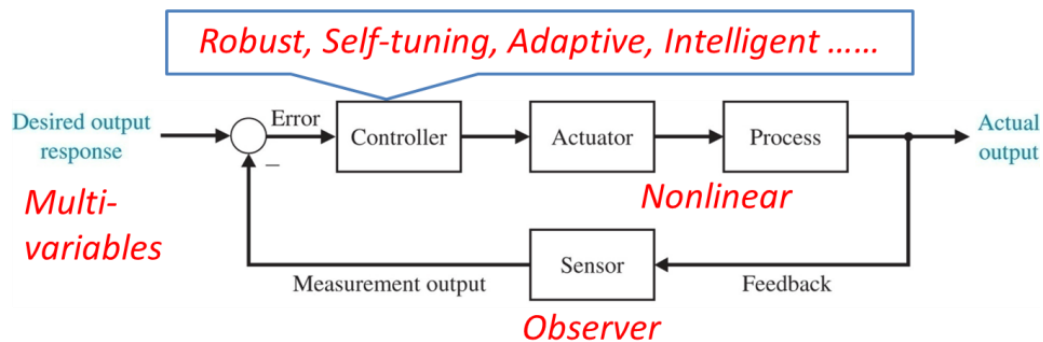

Figure 1: A typical $1^{st}$ and $2^{nd}$ generation feedback control system,
the modelling and design tools can be transfer function based and/or state-space based.

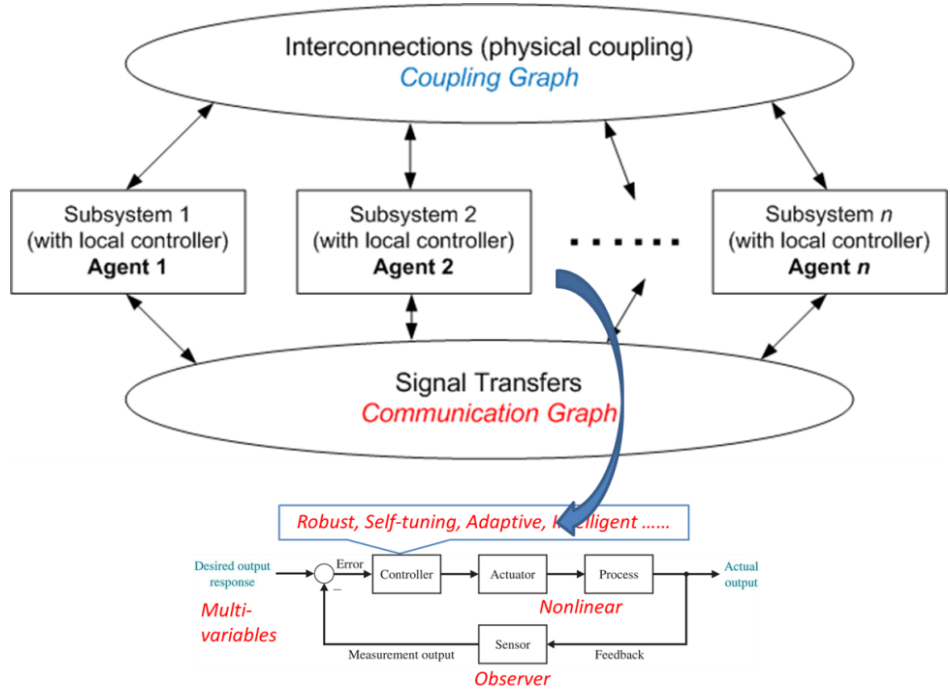

Figure 2: Control a network of subsystems, i.e., 3$^{rd}$ generation feedback control system

For the stability, Figure 3 is a typical representation of the stability in the 2$^{nd}$ generation of control theory, where "For any initial condition …… it will converge to the origin of the state space". On the other hand, Figure 4 is a typical representation of the stability in the 3$^{rd}$ generation of control theory, where in the case of formation, instead of $x(t) \to 0$, it is $\|x_i(t) - x_j(t)\| \to c$, where $c$ is a constant; and in the case of consensus, again instead of $x(t) \to 0$, it is $x_i(t) = x_j(t) \to f(\mathbf{x}(0))$. It will converge to a, normally non-zero, value which depends on the initial condition.

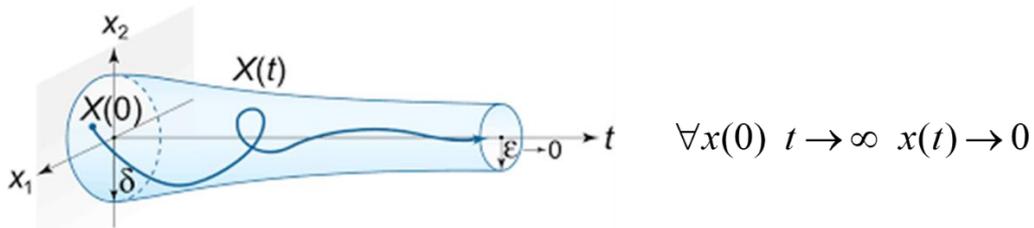

Figure 3: a typical representation of the stability in the 2$^{nd}$ generation of control theory

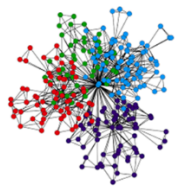

$$\forall_{i,j} \quad t \to \infty \quad \|x_i(t) - x_j(t)\| \to c \quad \text{Formation}$$
$$t \to \infty \quad x_i(t) = x_j(t) \to f(\mathbf{x}(0)) \quad \text{Consensus}$$

Figure 4: a typical representation of the stability in the 3$^{rd}$ generation of control theory

For the controllability, we do a basic overview as follows. Within the framework of the 2$^{nd}$ generation, it is from *A* and *B* matrixes to determine if a system is controllable. Under the NCS framework, it first addresses the *Structural Controllability*: i.e., quoted from [2]: "to determine the *minimum number D* of controllers required to ……". Then how to choose *D*, among all $N \gg D$, notes where control actions are to be applied to make an NCS fully controllable. Following the pioneering work of [2], there is a lot of interesting and active research. For example, see "2016 Outstanding Paper Award", titled: "Controllability metrics, limitations and algorithms for complex networks" [3]; a paper addressing controllability from the energy point of view, titled: "Physical controllability of complex networks"[4]; and two recent 2021 papers, titled "On the observability and controllability of large-scale IoT networks: reducing number of unmatched nodes via link addition"[5], and "Key nodes selection in controlling complex networks via convex optimization" [6] respectively.

Morden Artificial Intelligence (AI) is a thriving technology for many applications in various areas. Its impact on control is still not fully explored yet. Up to the 3$^{rd}$ generation of feedback control system design, "Desired output response" – as shown in the left of Figure 1 – is always given before controller design. Now this is changing. For example, for autonomous driving in very complicated environment, AI first works out the desired travelling trajectory as a reference signal, then the rest of control system to follow. Due to this and other AI's potential for control, we think that control theory and application has begun its 4$^{th}$ generation journey: "control in the new AI era".